\documentclass[12pt]{iopart}

\usepackage{iopams}
\usepackage{graphicx}

\expandafter\let\csname equation*\endcsname\relax
\expandafter\let\csname endequation*\endcsname\relax
\usepackage{amsmath}
\usepackage{color}

\newcommand{\nn}{\nonumber}
\newcommand{\bb}{\begin{equation}}
\newcommand{\ee}{\end{equation}}
\newcommand{\bbb}{\begin{align}}
\newcommand{\eee}{\end{align}}

\renewcommand{\eref}[1]{Eq.~(\ref{#1})}
\newcommand{\efig}[1]{Fig.~\ref{#1}}
\renewcommand{\Re}{{\rm Re}}

\newcommand{\Ar}{A}
\newcommand{\ff}{\makebox{\small $\frac{1}{2}$}}

\newcommand{\de}{\phi}
\newcommand{\ka}{k_{\alpha}}
\renewcommand{\e}{{\rm e}}
\newcommand{\drm}{\mathrm{d}}
\newcommand{\cb}{\textcolor{black}}

\begin{document}

\title{\cb{de Haas-van Alphen oscillations with non-parabolic dispersions}}
\author{Jean-Yves Fortin$^1$, Alain Audouard$^2$}

\address{$^1$Institut Jean Lamour, D\'epartement de Physique de la Mati\`ere
et des Mat\'eriaux, Groupe de Physique Statistique, CNRS UMR 7198,
BP 70239, F-54506 Vandoeuvre-l\`es-Nancy Cedex, France
\\ \ead{jean-yves.fortin@univ-lorraine.fr}
$^2$Laboratoire National des Champs Magn\'{e}tiques
Intenses (UPR 3228 CNRS, INSA, UGA, UPS) 143 avenue de Rangueil,
F-31400 Toulouse, France
\\ \ead{alain.audouard@lncmi.cnrs.fr}}

%


\begin{abstract}
de Haas-van Alphen oscillation spectrum of two-dimensional systems is
studied for general power law energy dispersion, yielding a Fermi surface area
of the form
$S(E)\propto E^\alpha$ for a given energy $E$. The case $\alpha=1$ stands for 
the parabolic energy dispersion. It is demonstrated that the periodicity of 
the magnetic oscillations in inverse field can depend notably on the
temperature. We evaluated analytically the Fourier
spectrum of these oscillations to evidence the frequency shift
and smearing of the main peak structure as the temperature increases.
\end{abstract}

\pacs{71.10.Ay, 71.18.+y, 73.22.Pr}

\maketitle
\section{Introduction}
Due to their parabolic energy dispersion, most metallic systems, including
organic conductors, heavy fermions, give way to de Haas-van Alphen (dHvA)
oscillations periodic in inverse magnetic field $B$, with a periodicity
proportional to the Fermi surface area supporting the cyclotronic trajectory
of the quasiparticle. When only one band is involved, the Fourier spectrum is
composed of a series of peaks with a fundamental frequency and its harmonics.
The precise amplitude of these peaks is given by the Lifshitz-Kosevich
(LK) formula~\cite{Abrikosov,Ziman,LK2bis,LK2,Roth:1966}, originally derived for three
dimensional metals with parabolic band structures (for which the Fermi surface
area grows or decreases linearly with energy). It also holds for strongly
two-dimensional band structures where metallic layers are separated from each
other by insulating layers, as it is the case of the organic metals
$\theta$-(ET)$_4$CoBr$_4$(C$_6$H$_4$Cl$_2$)~\cite{Au12} or
$\theta$-(ET)$_4$ZnBr$_4$(C$_6$H$_4$Cl$_2$)~\cite{fortin:2015jpc}, albeit 
oscillation of the chemical potential in magnetic field must be taken into 
account for these non-compensated metals.

In contrast, deviation to parabolic energy dispersion, namely in
presence of a curvature $S''(E)\neq 0$ when the Fermi surface area grows
non-linearly with the energy, leads to an additional Onsager
phase in the oscillations that is dependent on the magnetic field $B$ and the
temperature
$T$. In the limit where the ratio $B/T$ is either large or small, this phase is
proportional to $T^2/B$ ~\cite{fortin:2015epjb}, which yields, 
in the considered field ranges, a temperature dependent frequency. Notably, this
behavior should be observed in the case of Dirac fermions, the energy
dispersion of which is linear with the momentum~\cite{mcclure66} and liable to
yield non-parabolic deviations. Indeed, the energy curvature of the
Fermi surface area is proportional in this case to
$S''(E)=2\pi/(h v_F)^2$ where $v_F\simeq 10^6$m/s is the Fermi velocity.

It must be pointed out that the observed frequency changes and (or) phase
shifts of the oscillations can also be attributed to other physical effects
which have to be treated separately. For example, spin-orbit coupling leads to
a splitting of the Fermi surface, hence of the dHvA frequency with a magnitude
proportional to $B^2$ and to the effective mass~\cite{Mineev:2005}. Such
splitting has been considered in data relevant to e.g.
bilayer underdoped high-T$_c$ cuprates~\cite{Sebastian:2014}.
Phase shift is observed in the presence of
magnetic breakdown when the Fermi surface is composed of several bands between
which the quasiparticles can tunnel, leading to giant orbits. In this
case, the field-dependent phase is linked to the probability of
tunneling~\cite{Fa66,Sl67,Slutskin:1968}. It depends on the ratio between the
breakdown field and the field itself, but not the temperature. Its
origin comes from a linearization of the Fermi surface near the tunneling region
for which the quantum wavefunctions can be solved. This problem is similar to
the Zener effect~\cite{Vitanov:2008} but in a magnetic field. Such phase
has been observed and studied in the organic compound
$\theta$-(BEDT-TTF)$_4$CoBr$_4$(C$_6$H$_4$Cl$_2$)
where the breakdown field is close to 35 T~\cite{2013SyntM}. To end
with examples, strong deviation from parabolicity is observed within the
tight-binding model in two dimensions where the band gap closes at
half-filling, yielding non LK behavior of the oscillation
amplitude~\cite{Yong:1996}.

In this paper, we focus on special
cases of power law spectrum dependence of the Fermi surface area $S(E)\propto
E^{\alpha}>0$, where the exponent $\alpha$ is taken as a control parameter,
either negative or positive, standing respectively for hole or electron-type
quasiparticles. In section~\ref{section_GP}, magnetization is 
evaluated from an exact resummation of the grand
potential in the quasi-classical limit of small fields, and the result is
given by~\eref{Ip}, which is a general formula from which dHvA oscillations can 
be analyzed.
In section~\ref{section_powerlaw}, we apply our analysis to power law 
dispersion, where exact integrals can be performed. Both cases $\alpha<0$ and
$\alpha>0$ are studied and the peak structure of the temperature-dependent
Fourier transform is discussed. In particular, calculations are made
for $\alpha<0$ and $\alpha\ge 1$, relevant to hole-type
quasiparticles and electron-type Dirac fermions, respectively. In the last
section~\ref{section_LT}, we provide a low-temperature expression of
magnetization in the general case.

\section{Expansion of the grand potential\label{section_GP}}
Let us consider a band structure in a two-dimensional system, yielding a
closed surface area in the Brillouin zone equal to $S(E)=\oint_{E}k_y\drm
k_x/4\pi^2$ at constant energy $E$. In the
parabolic case $S(E)=E$ and for Dirac fermions $S(E)=E^2$. The quantization of
Landau levels is given by the Bohr-Sommerfeld rule $S(E_n)=b(n+\gamma)$, with
the reduced field $b=eB/h$ in inverse area units where $-e$ is the electron
charge. Indeed, for typical elementary cell areas, $b$ is small 
and a semi-classical analysis can be applied. We want to study in the next 
section~\ref{section_powerlaw} the oscillations
of the magnetization for the class of surfaces with a power law dependence
$S(E)=E^{\alpha}$, where $\alpha=1$ stands for a parabolic surface.

To be more rigorous with the units, we should consider the
physical coefficient of proportionality $\ka$ between $S(E)$ and $E^{\alpha}$
(with units $[\ka]=L^{-2}E^{-\alpha}$ where $L$ is a length and $E$ an energy)
but for simplicity we will ignore it. It can be restored at the end of the computations
by rescaling $b\rightarrow b/\ka$ (with $[b]=L^{-2}$ and $[b/\ka]=E^{\alpha}$), and
frequency $F\rightarrow F/\ka$ (with $[F]=L^{-2}$ and $[F/\ka]=E^{\alpha}$)
accordingly. For example, the surface areas for the
free fermions and Dirac fermions in real units are respectively given by
\bb\label{SE}
S(E)=m^*E/(2\pi\hbar^2),\;S(E)=\pi E^2/(2\pi\hbar v_F)^2
\ee
where $m^*$ is the effective mass of the quasiparticle. The temperature $T$ 
will be expressed in units of $\mu/k_B$ in the numerical applications further 
below, where $\mu$ is the chemical potential. More specifically, the 
fundamental frequency $F_0$ of
the oscillation spectrum is equal to 
$F_0=hS(\mu)/e=(h/e)k_{\alpha}\mu^{\alpha}$ in Tesla, since $h/e$ is a quantum 
flux and $S(\mu)$ is the area of the
cyclotronic trajectory in the Brillouin zone (inverse square length). The 
reduced temperature $t$ can then be expressed as
$t=k_BT/\mu=k_B(k_{\alpha}h/e)^{1/\alpha}T/F_0^{1/\alpha}$, with $T$ expressed 
in Kelvin and $F_0$ in Tesla. For $\alpha=1$, 
one finds that $t=0.745 T (m^*/m_e)/F_0$, and when $\alpha=2$, one has instead 
$t=2375.2 T/(v_FF_0^{1/2})$.

In this section, we will give the expression of the oscillatory part of the grand potential in the more
general case, not restricted to the power law class, and the main result is given by~\eref{Ip}.
We first consider the oscillating part of the grand potential obtained from the Poisson formula for
an arbitrary discrete Landau spectrum~\cite{LandauVol9}
\begin{align}\nn
\frac{\Omega_{osc}}{\Ar}=&-\frac{2b}{\beta}
\Re
\sum_{p=1}^{\infty}\int_0^{\infty}\drm n\log
\left [
1+\e^{\beta(\mu-E_n)} \right ]\e^{2i\pi p n}
\\ \label{Omega_0}
=&-\frac{2b}{\beta}\Re
\sum_{p=1}^{\infty} I_p
\end{align}
where $\beta=(k_BT)^{-1}$ and $\Ar$ is the sample area. Following
the analysis leading to the LK
result, we perform an integration by parts of the partial quantities $I_p$
\begin{align}\nn
I_p=&-\frac{1}{2i\pi p}\log[1+\e^{\beta(\mu-E_0)}]-\frac{\beta}{(2i\pi
p)^2}E_0'\de_0
\\ \label{Ip0}
&+\frac{\beta^2}{(2i\pi p)^2}
\int_0^{\infty}\drm n\e^{2i\pi p n}E_n'^2\de_n'
-\frac{\beta}{(2i\pi p)^2}
\int_0^{\infty}\drm n\e^{2i\pi p n}E_n''\de_n
\end{align}
where we have defined $\de_n=\de(\beta[\mu-E_n])$ and
$\de(x)=(1+\e^{-x})^{-1}$. In the LK theory, i.e. for a parabolic band, the
last term is zero since $E_n''=0$. And the
function $\beta\de_n'$ behaves like a Dirac distribution in the low temperature
limit around the Fermi energy~\footnote{Indeed the function
$\beta\de'(\beta[\mu-E])\simeq \delta(E-\mu)$ when $T\rightarrow 0$.}. The
LK approximation is relying on computing this integral in the complex plane
in the low temperature limit when $-\beta\mu\rightarrow -\infty$, after a
change of variable $n(E)\rightarrow E$
is performed. The poles of the function $\de'$ considered for applying the
residue theorem are located on the upper plane. After performing a resummation
over the poles as a geometrical series, one obtains an expression for the
thermal amplitude that decays exponentially with the temperature. The first
term on the right hand side of~\eref{Ip0} is a pure imaginary number, and
despite the divergence of the sum over $p$, it does not contribute to the real
part of $I_p$
and can be discarded in the following. In the general non-parabolic case, we
can perform an additional integration by parts on the last term in order to
single out the function $\beta\de_n'$
\begin{align}\nn
I_p=&-\frac{\beta}{(2i\pi
p)^2}E_0'\de_0+\frac{\beta}{(2i\pi
p)^3}E_0''\de_0
+\frac{\beta^2}{(2i\pi p)^2}
\int_0^{\infty}\drm n\e^{2i\pi p n}E_n'^2\de_n'
\\ 
-&\frac{\beta^2}{(2i\pi p)^3}
\int_0^{\infty}\drm n\e^{2i\pi p n}E_n'E_n''\de_n' 
+\frac{\beta}{(2i\pi p)^3}
\int_0^{\infty}\drm n\e^{2i\pi p n}E_n'''\de_n
\end{align}
The last integral whose integrand is proportional to $\de_n$ can be furthermore
integrated by parts, and this process can be repeated indefinitely. One obtains
a formal series as the result of all the consecutive partial integrations which
is expressed as
\begin{align}
I_p=&\beta\sum_{k=1}^{\infty}\frac{(-1)^k}{(2i\pi
p)^{k+1}}E_0^{(k)}\de_0
+\beta^2
\int_0^{\infty}\drm n\e^{2i\pi p n}E_n'\de_n'
\sum_{k=1}^{\infty}\frac{(-1)^{k+1}}{(2i\pi
p)^{k+1}}E_n^{(k)}
\end{align}
This series can be resummed and put into a single integral using the formula
$\sum_{k=0}^{\infty}(-x)^kf^{(k)}(n)=\int_0^{\infty}\drm u\e^{-u}f(n-ux)$. One
easily obtains a compact form for the partial quantities $I_p$ as the sum
of two terms
\begin{align}\nn
I_p=&\frac{\beta\de_0}{2i\pi p}\int_0^{\infty}\drm u\e^{-u}
\left [
E_{-u/(2i\pi p)}-E_0 \right ]
\\ \label{Ip}
-\frac{\beta^2}{2i\pi p}&
\int_0^{\infty}\drm n\e^{2i\pi p n}E_n'\de_n'
\int_0^{\infty}\drm u\e^{-u}
\left [
E_{n-u/(2i\pi p)}-E_n \right ]
\end{align}
This expression encompasses all the deviations from parabolicity without
linearization and is the main formula of this work from which we can deduce
physical implications of any specific dispersion $E_n$ on the dHvA oscillatory 
Fourier spectrum.
The simplest example is provided by the parabolic dispersion,
$E_n=b(n+\gamma)$ with $\gamma=1/2$, for which the integral over $u$ in~\eref{Ip} can 
be easily performed. In this case, it is convenient to change the variable 
$n\rightarrow n(E)=E/b-\gamma$ to obtain
\bb\nn
I_p=-\frac{\beta b\de_0}{(2i\pi p)^2}
+\frac{\beta^2b}{(2i\pi p)^2}
\int_0^{\infty}\drm E\e^{2i\pi p(E/b-\gamma)}\de'(\beta[\mu-E])
\ee
Then the integral of the previous expression can be computed after the
rescaling of the energy $2\pi x=\beta(E-\mu)$
\begin{align}
&\int_0^{\infty}\drm E\e^{2i\pi p(E/b-\gamma)}\de'(\beta[\mu-E])
=
\e^{2i\pi p\mu/b-2i\pi p\gamma}\int_{-\beta\mu/2\pi}^{\infty}\frac{\pi
\drm x}{2\beta}\frac{\e^{4i\pi^2 px/(\beta b)}}{\cosh^2(\pi x)}
\end{align}
At low temperature, it is usual to replace the lower bound $-\beta\mu/2\pi$
by $-\infty$. Then the residue theorem can be applied in the upper plane, where
the zeroes of function $\cosh(\pi x)$, which are the poles of the integrand, are
given by $x_n=i(n+1/2)$, $n\ge 0$. After the resummation over the $x_n$, one
obtains a temperature dependent damping amplitude $R$ defined by
\bb
\int_{-\infty}^{\infty}\frac{\pi
\drm x}{2}\frac{\e^{4i\pi^2 px/(\beta b)}}{\cosh^2(\pi
x)}=R\left (\frac{2\pi^2 p}{\beta b} \right )
,\;R(x)=\frac{x}{\sinh(x)}
\ee
Then, the dominant part of the grand potential is given by
\bb
\frac{\Omega_{osc}}{\Ar}\simeq
\sum_{p=1}^{\infty}
\frac{b^2}{2\pi^2p^2}R\left (\frac{2\pi^2 p}{\beta b} \right )\cos\left (
2\pi p \frac{\mu}{b}-2\pi p\gamma \right )
\ee
The oscillating part of the magnetization is defined
by $m_{osc}=-\Ar^{-1}\partial
\Omega_{osc}/\partial b$, or, with a good approximation
\bb\label{mosc_para}
m_{osc}(b)\simeq
-\sum_{p=1}^{\infty}
\frac{\mu}{\pi p}R\left (\frac{2\pi^2 p}{\beta b} \right )\sin\left (
2\pi p \frac{\mu}{b}-2\pi p\gamma \right )
\ee
which is the LK formula. After restoring the correct units $b\rightarrow b/k_1=2\pi\hbar^2
b/m^*=\hbar\omega_c$, with cyclotron
frequency $\omega_c=eB/m^*$, one finds that the fundamental frequency is equal 
to $F_0=m^*\mu/(\hbar e)$. Now the thermal damping amplitude
$R$ is usually written as $R(2\pi^2 pk_BT/\hbar\omega_c)=R(u_0(m^*/m_e)pT/B)$,
with
$u_0=2\pi^2k_Bm_e/\hbar e=14.7$ T/K for an electron of mass $m_e$. This thermal
factor is essential in determining the quasi-particle mass $m^*$ from the 
temperature dependence of the Fourier amplitude of
each harmonic $pF_0$.
\section{Surfaces with a power law dependence\label{section_powerlaw}}
In this section we consider the general case of a power law dependence of the area
$S(E)=E^{\alpha}$. When $\alpha$ is negative the surface is a hole-type
surface, with a negative geometrical mass $S'(E)<0$,  whereas when $\alpha>0$
the surface is an electronic band with a positive mass. \eref{Ip} can be
rewritten, after discarding the first term
which is assumed to be small, as
\begin{align}\nn
I_p=&-\frac{\beta^2}{2i\pi p}\int_0^{\infty}\drm EE\e^{2i\pi
p(E^{\alpha}/b-\gamma)}
\de'(\beta[\mu-E])
\\
\times &\int_0^{\infty}\drm u\e^{-u}\left [
\left (1-\frac{bu}{2i\pi pE^{\alpha}}\right )^{1/\alpha}-1\right ]
\end{align}
Magnetization, defined by $m_{osc}=-\Ar^{-1}\partial
\Omega_{osc}/\partial b$, is dominated by the derivative of the
exponential phase $\e^{2i\pi pE^{\alpha}/b}$, which gives a main contribution in
$1/b^2$, and after a change of variable $bu/pE^{\alpha} \rightarrow u$, one
obtains the expression
\begin{align}\nn
&m_{osc}(b)\simeq \frac{2}{b^2}\Re \sum_{p\ge
1}p\e^{-2i\pi p\gamma}\int_0^{\infty}\drm
EE^{1+2\alpha}\e^{2i\pi pE^{\alpha}/b}
\\ &\times\beta
\de'(\beta[\mu-E]) \label{mag_gen}
\int_0^{\infty}\drm u\e^{-pE^{\alpha}u/b}\left [
\left (1-\frac{u}{2i\pi}\right )^{1/\alpha}-1\right ]
\end{align}
In the parabolic case $\alpha=1$, one easily finds
\bb
m_{osc}(b)\simeq -2\Re \sum_{p\ge
1}\frac{1}{2i\pi p}\int_0^{\infty}\drm
E E\e^{2i\pi p(E/b-\gamma)}\beta
\de'(\beta[\mu-E])
\ee
We want to study the general Fourier spectrum  $A(F)$ of the magnetization with
respect to the inverse field $x=1/b$, where $x$ is taken by extension from $-\infty$ to
$+\infty$, and from which $m_{osc}(b)=\Re
\int_{-\infty}^{\infty}\drm F A(F)\e^{2i\pi F/b}$. In the parabolic case the
amplitude can be computed exactly since the integral over $x$ involves only a
Dirac function
\begin{align}\nn
A(F)= &-\sum_{p\ge
1}\frac{1}{i\pi p}\int_{-\infty}^{\infty}\drm x\int_0^{\infty}\drm
E E\e^{2i\pi (pE-F)x-2i\pi p\gamma}
\beta\de'(\beta[\mu-E])
\\ \label{AF_para}
=& -2F\sum_{p\ge
1}\frac{e^{-2i\pi p\gamma}}{2i\pi p^3}\beta
\de'(\beta[\mu-F/p])\theta(F)
\end{align}
Here $\theta$ is the Heaviside function. The amplitude $A(F)$ is peaked at
frequencies $F=p\mu$ as expected. $A(F)$ is
purely imaginary in the case where $\gamma=1/2$. In the LK 
theory,
the factor $F$ in front of~\eref{AF_para} is replaced by the saddle point solution $p\mu$.

\subsection{Hole-type quasiparticles ($\alpha<0$)}

\begin{figure*}[ht]
\centering
\includegraphics[width=0.45\columnwidth,clip,angle=0]{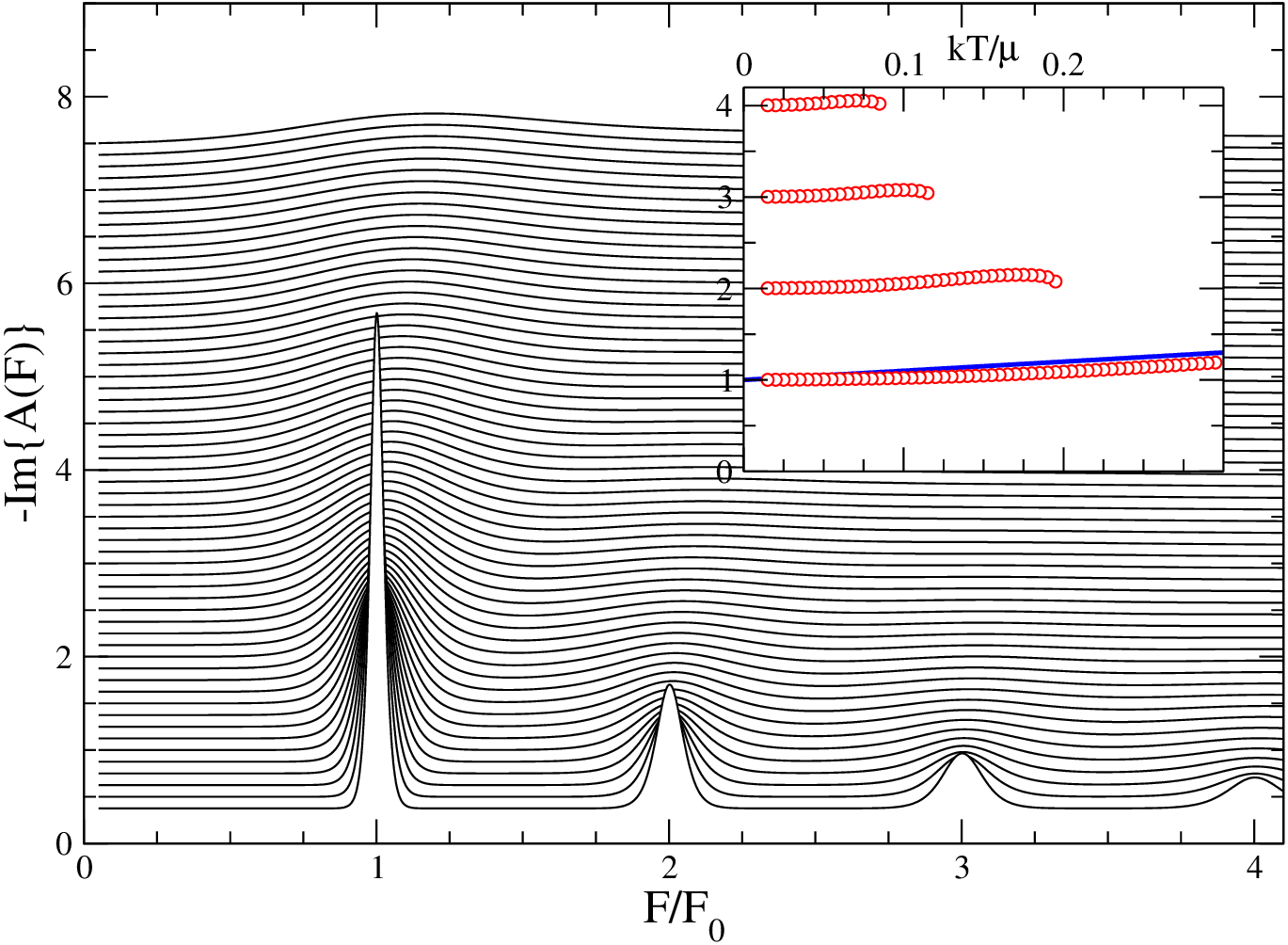}(a)
\includegraphics[width=0.45\columnwidth,clip,angle=0]{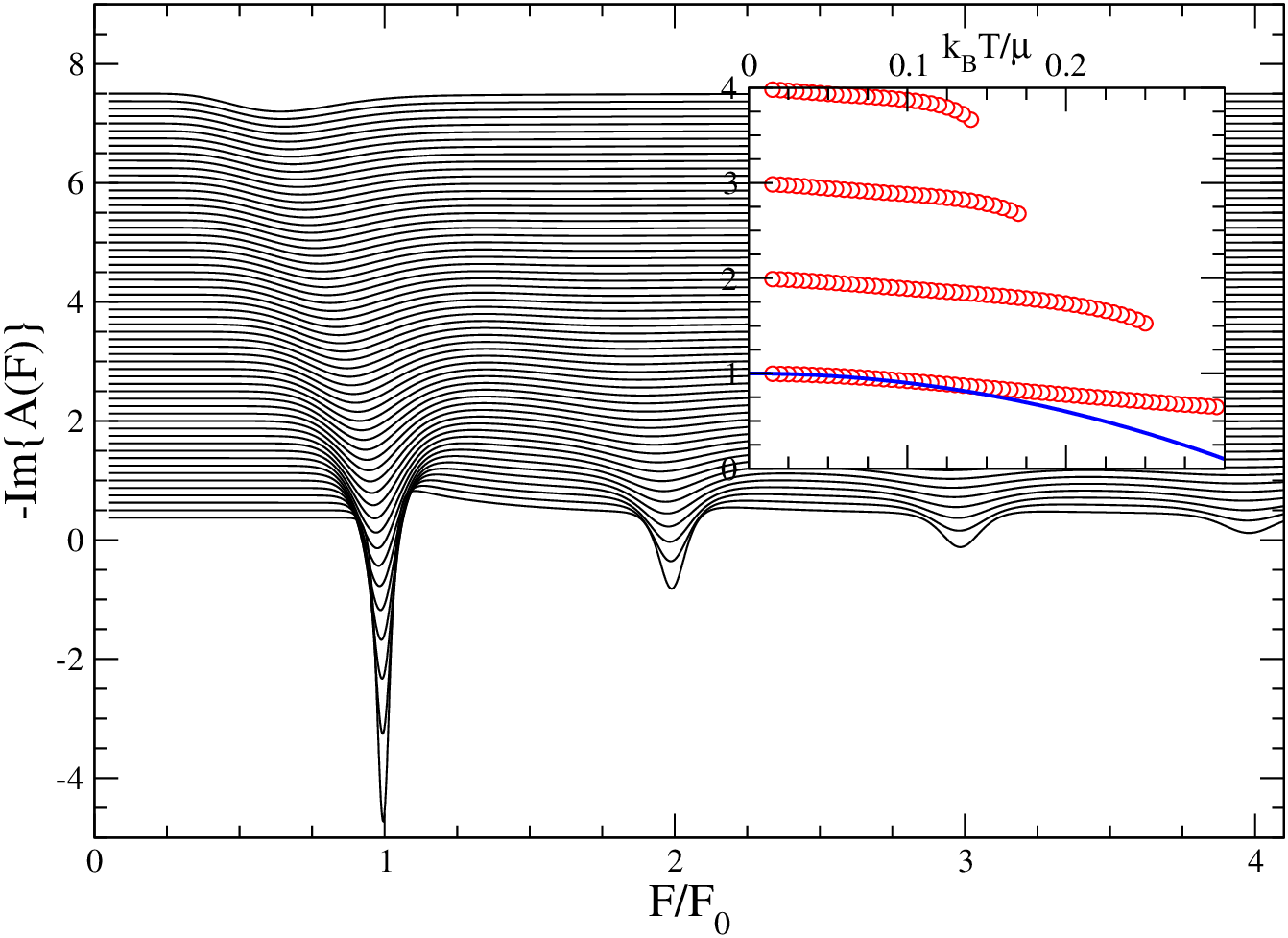}(b)
\caption{Plot of the Fourier amplitude $-\Im\{A(F)\}$ for the cases
$\alpha=1$ (a) and $\alpha=-1$ (b), respectively, and for
temperatures ranging from t=0.01 up to t=0.3, where  $t=k_BT/\mu$. Here we have taken $\gamma=0$ in both
cases for simplicity, otherwise the peaks take alternative signs. 
In the insets is represented the temperature dependence of the extrema. For 
electron-type particles, the peaks are moving to the right as the temperature
increases, whereas for hole-type they move to the opposite 
direction. Solid blue lines are the
low temperature approximation of the first harmonics given 
by~\eref{F0T}. Curves are obtained from the numerical resolution 
of~\eref{AF_pos} and~\eref{AF_neg} respectively, and the positions of the 
extrema are found numerically.}
\label{fig1}
\end{figure*}

In this section, we focus on the case $\alpha<0$ with the aim to evaluate the
Fourier transform of the magnetization~\eref{mag_gen}. It is useful to
consider the following formula representation of the last integral over $u$
in~\eref{mag_gen} when $\alpha<0$ for complex $z$ with positive real and
imaginary parts
\begin{align}\nn
\int_0^{\infty}\drm u\e^{-zu}\left [(1-\frac{u}{2i\pi})^{1/\alpha}-1\right
]=&-z^{-1}-2i\pi \e^{-2i\pi z}\int_1^{\infty}\drm uu^{1/\alpha}\e^{2i\pi
z u}
\\ \label{exp_int}
=&-z^{-1}-2i\pi \e^{-2i\pi z}E_{-1/\alpha}(-2i\pi z)
\end{align}
where $E_n(z)=\int_1^{\infty}\drm u u^{-n}\e^{-zu}$ are the exponential
integral.
This representation is only valid for $\alpha<0$ otherwise the integral
on the right hand side would diverge at large argument.
Magnetization can then be written as
\begin{align}\nn
m_{osc}(b)
=&-\frac{2}{b}\Re \sum_{p\ge
1}\e^{-2i\pi p\gamma}\int_0^{\infty}\drm EE^{1+\alpha}\beta
\de'(\beta[\mu-E])
\\ \nn
&\times \left [\e^{2i\pi pE^{\alpha}/b}+\frac{2i\pi pE^{\alpha}}{b}
\int_1^{\infty}\drm uu^{1/\alpha}\e^{2i\pi pE^{\alpha}u/b
}
\right ]
\\ \label{mosc_alpha_neg}
=&-\frac{2}{b}\Re \sum_{p\ge
1}\e^{-2i\pi p\gamma}\int_0^{\infty}\drm EE^{1+\alpha}\beta
\de'(\beta[\mu-E])
\\ \nn
&\times \left [\e^{2i\pi pE^{\alpha}/b}+\frac{2i\pi pE^{\alpha}}{b}
E_{-1/\alpha}(-2i\pi pE^{\alpha}/b)
\right ]
\end{align}
The integral over $E$ is convergent near the origin only if $-1<\alpha<0$ due to the presence
of a term proportional to $E^{1+2\alpha}$ in the expression. Otherwise a cut-off should be
introduced when $E$ is small, for which the Fermi surface area 
$S(E)$ diverges strongly. The Fourier amplitude $A(F)$ can now be evaluated 
from~\eref{mosc_alpha_neg}
after performing different integrations over the variables $x=1/b$ and $u$
respectively
\begin{align}\label{AF_neg}
A(F)=
\frac{i}{\pi\alpha}\frac{\partial}{\partial F}\sum_{p\ge
1}\frac{\e^{-2i\pi p\gamma}}{p^{1/\alpha}}
F^{1/\alpha-1}
\int_{(F/p)^{1/\alpha}}^{\infty}\drm E E^{\alpha}
\beta\de'(\beta[\mu-E])
\theta(F)
\end{align}
$A(F)$ is a pure imaginary number when $\gamma=1/2$ or $\gamma=0$.
We have plotted the imaginary part $-\Im A(F)$ in~\efig{fig1}(b), for several values of
temperatures, and compared it with the parabolic
case~\efig{fig1}(a) from~\eref{AF_para}. The dominant peak structure is
centered at harmonics $F=p/\mu$ but, contrary to the parabolic case, the
peaks are moving to lower values as the temperature increases. In the parabolic
case the deviation is present although less pronounced. It is
due to the presence of the factor $F$ in front of~\eref{AF_para}, which is 
considered as constant and equal to $\mu$ in the LK theory. 
But the deviation is too small to be observed at temperatures where the 
amplitude is not negligibly small. In any case, this approximation is valid for 
the temperature range explored in experiments.
In short, as the temperature increases, the frequency of the Fourier components 
decreases and increases, for electron and hole type quasiparticles, 
respectively, the frequency variation being more pronounced
for hole type quasiparticles.
%
\begin{figure*}[ht]
\centering
\includegraphics[width=0.45\columnwidth,clip,angle=0]{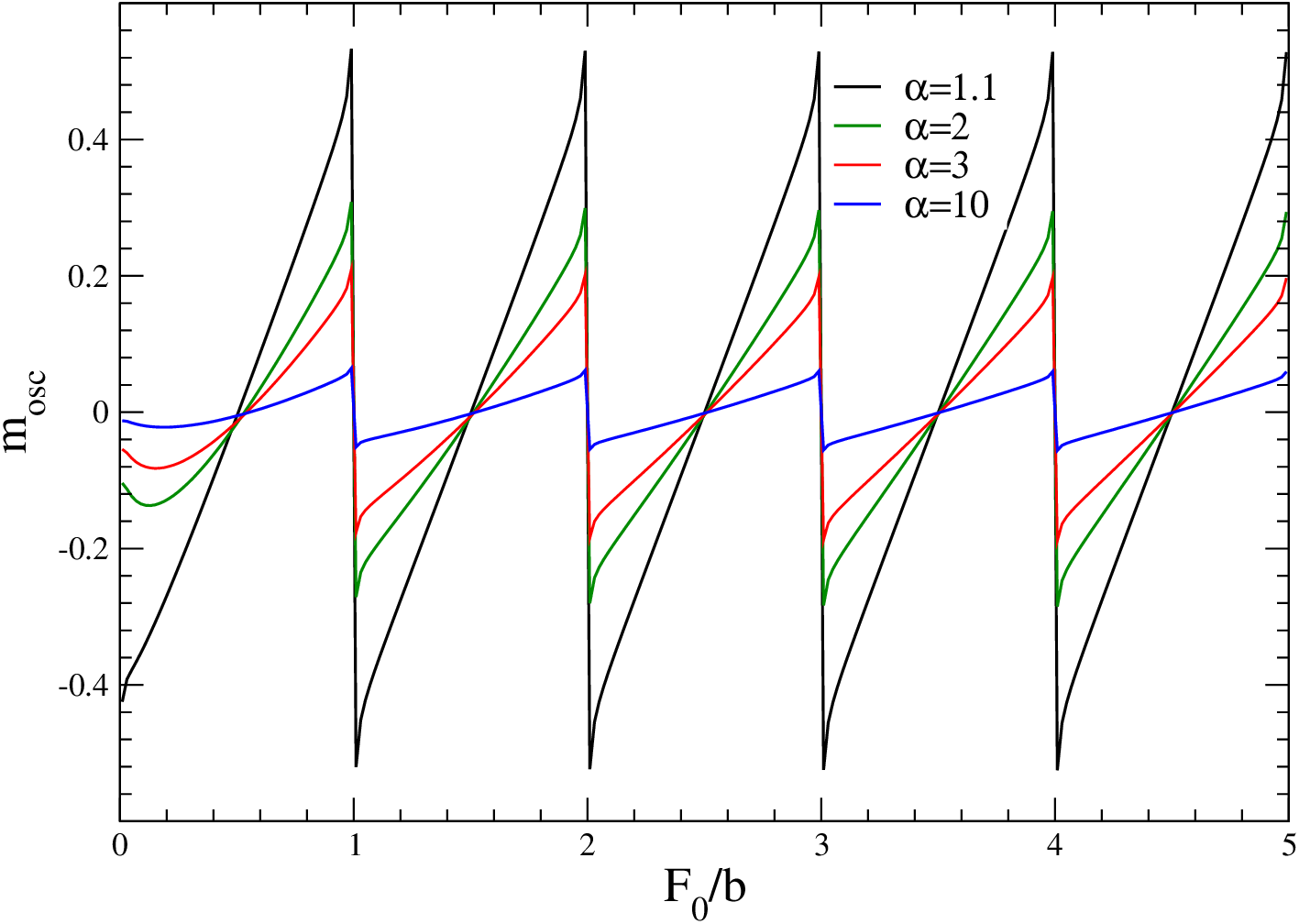}(a)
\includegraphics[width=0.45\columnwidth,clip,angle=0]{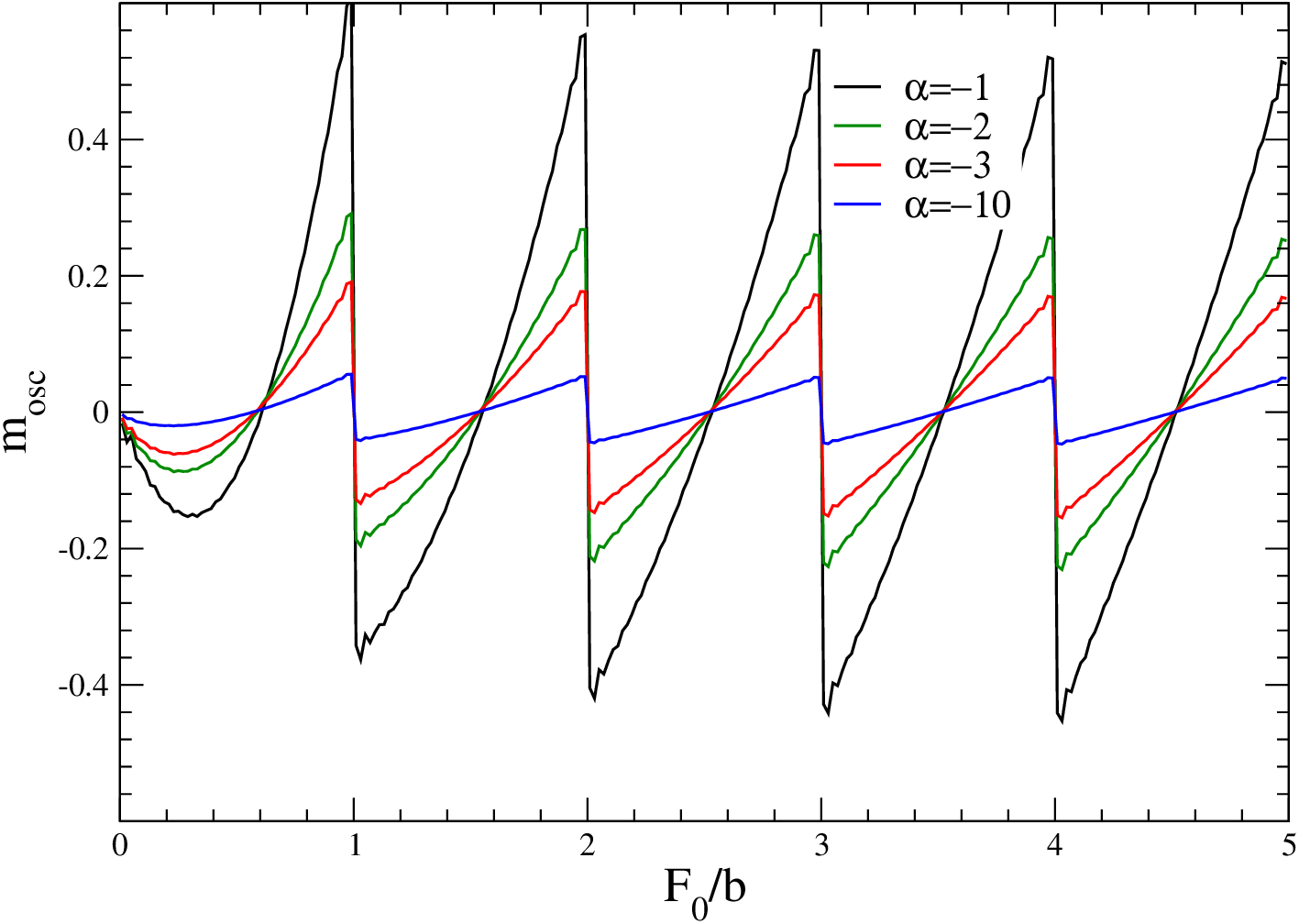}(b)
\caption{Plot of the magnetization at zero temperature for
different values of $\alpha$ as a function of the inverse field $F_0/b$.}
\label{fig2}
\end{figure*}

\subsection{Electron-type quasiparticles ($\alpha>1$)}

In this section, we consider the Fourier amplitude of~\eref{mag_gen} when
$\alpha\ge 1$. For other cases $0<\alpha<1$ the techniques developed below
can be applied as well with some modifications. We would like to get a 
representation of the last integral on the right hand side of~\eref{mag_gen} 
still in terms of exponential integrals such as in~\eref{exp_int}. 
Indeed, the exponential integral on the right hand side of~\eref{exp_int} is 
not finite for imaginary argument and $\alpha>0$ as seen before. However we can 
use a differentiation
\begin{align}
\int_0^{\infty}\drm u\e^{-zu}\left (1-\frac{u}{2i\pi}\right )^{1/\alpha}
=\frac{\e^{-2i\pi z}}{2i\pi}\frac{\partial}{\partial z}\left [
\int_0^{\infty}\drm u\e^{-z(u-2i\pi)}\left (1-\frac{u}{2i\pi}\right
)^{1/\alpha-1}\right ]
\end{align}
Since $1/\alpha-1$ is negative for $\alpha>1$, we can apply~\eref{exp_int} to
have a useful representation for the integral after the differentiation
operator. In particular, we can write the last integral of~\eref{mag_gen} as
\begin{align} \nn
\int_0^{\infty}\drm u\e^{-pE^{\alpha}u/b}
\left (1-\frac{u}{2i\pi}\right )^{1/\alpha}
=&\frac{b\e^{-2i\pi pE^{\alpha}/b}}{2i\pi pE^{\alpha}}
\frac{\partial}{\partial\lambda}
\int_0^{\infty}\drm u\e^{-pE^{\alpha}(u-2i\pi)/b}
\left (1-\frac{u}{2i\pi}\right )^{1/\alpha-1}\Big |_{\lambda=1}
\\ \nn
=&-\frac{b\e^{-2i\pi pE^{\alpha}/b}}{pE^{\alpha}}
\frac{\partial}{\partial\lambda}
\int_1^{\infty}\frac{\drm u}{u^{1-1/\alpha}}\e^{2i\pi pE^{\alpha}\lambda
u/b}\Big |_{\lambda=1}
\\ \nn
=&-\frac{b\e^{-2i\pi pE^{\alpha}/b}}{pE^{\alpha}}
\frac{\partial}{\partial\lambda}
\int_{\lambda}^{\infty}\frac{\lambda^{-1/\alpha}\drm u}{u^{1-1/\alpha}}\e^{2i\pi
pE^{\alpha}
u/b}\Big |_{\lambda=1}
\\
=&\frac{b\e^{-2i\pi pE^{\alpha}/b}}{pE^{\alpha}}
\left [\e^{2i\pi pE^{\alpha}/b}+\frac{1}{\alpha}\int_{1}^{\infty}\frac{\drm 
u}{u^{1-1/\alpha}}\e^{2i\pi pE^{\alpha}u/b} \right ]
\end{align}
From this representation, one can express the magnetization as
\begin{align}\label{mag_alpha_pos}
m_{osc}(b)=&
\frac{2}{\alpha b}\Re \sum_{p\ge
1}
\e^{-2i\pi p\gamma}\int_0^{\infty}\drm E E^{1+\alpha}\beta
\de'(\beta[\mu-E])
\int_1^{\infty}\frac{\drm u\e^{2i\pi pE^{\alpha}
u/b}}{u^{1-1/\alpha}}
\end{align}
Then we can write the Fourier amplitude of~\eref{mag_gen} as
\begin{align}\nn
A(F)=&
\frac{2}{\alpha}\sum_{p\ge
1}\int_{-\infty}^{\infty}\drm x x
\e^{-2i\pi p\gamma-2i\pi Fx}
\int_0^{\infty}\drm E E^{1+\alpha}\beta
\de'(\beta[\mu-E])
\int_1^{\infty}\frac{\drm u\e^{2i\pi pE^{\alpha}
ux}}{u^{1-1/\alpha}}
\\ 
=&\frac{i}{\pi\alpha}\frac{\partial}{\partial F}\sum_{p\ge
1}\e^{-2i\pi p\gamma}\int_0^{\infty}\drm E E^{1+\alpha}\beta
\de'(\beta[\mu-E])
\int_1^{\infty}\frac{\drm u}{u^{1-1/\alpha}}
\delta(F-pE^{\alpha}u)
\end{align}
After integration over $u$ one finally obtains the simple
expression
\begin{align}\label{AF_pos}
A(F)=&\frac{i}{\pi\alpha}\frac{\partial}{\partial F}\sum_{p\ge
1}\frac{\e^{-2i\pi p\gamma}}{p^{1/\alpha}}
F^{1/\alpha-1}
\int_0^{(F/p)^{1/\alpha}}\drm E E^{\alpha}\beta
\de'(\beta[\mu-E])\theta(F)
\end{align}
This expression is the same as~\eref{AF_neg} except for the interval integration. We recover in particular the
result~\eref{AF_para} in the parabolic case $\alpha=1$ by simple differentiation
of the integral since the term $F^{1/\alpha-1}=1$. For Dirac fermions,
$\alpha=2$ and $\gamma=0$, we can check the formula~\eref{AF_pos} with direct
computation of the Fourier transform of~\eref{mag_gen}:
\begin{align}\nn
A(F)=&\frac{1+i}{2}\sum_{p\ge 1}\frac{1}{\sqrt{p}}\int_0^{\infty}\drm E E^2
\beta\de'(\beta[\mu-E])
\\ \times&
\int_{-\infty}^{\infty}\drm x \sqrt{x}
\e^{-2i\pi Fx}{\rm erfc}\left (\sqrt{\pi pE^2x}(1-i) \right )
\end{align}
The integration over $x$ seems not easy to perform at first, but the erfc
function is related to the integral function of argument $1/2$ by ${\rm
erfc}(z)=z\pi^{-1/2}E_{1/2}(z^2)$
\begin{align}
A(F)=&\sum_{p\ge 1}\int_0^{\infty}\drm E E^3
\beta\de'(\beta[\mu-E])
\int_{-\infty}^{\infty}\drm x x\e^{-2i\pi Fx}E_{-1/2}(-2i\pi pE^2 x)
\end{align}
Then we use the integral representation of function $E_{1/2}$ which allows
to perform the integration over $x$. This leads to the same result
as~\eref{AF_pos} given by
\begin{align}
A(F)=&\frac{1}{4i\pi}\sum_{p\ge 1}\frac{\theta(F)}{\sqrt{p}}\Big [
F^{-3/2}\int_0^{\sqrt{F/p}}\drm E E^2
\beta\de'(\beta[\mu-E])
-p^{-3/2}\beta\de'(\beta[\mu-\sqrt{F/p}]) \Big ]
\end{align}
The amplitudes $A(F)$ in~\eref{AF_neg} and~\eref{AF_pos} are both pure imaginary,
and the magnetization is
given by $m_{osc}(b)=-\int_0^{\infty}\Im\{A(F)\}\sin(2\pi F/b)$.
In~\efig{fig2} we have represented the magnetization for different values of
$\alpha$, positive and negative respectively. As $\alpha$ increases, the
amplitude of the oscillations decreases. This is due to the $\alpha$-dependence
of the effective mass that we defined by
\bb\label{eff_mass}
m^*=2\pi\hbar^2\partial S(\mu)/\partial \mu=
2\pi\hbar^2k_{\alpha}\alpha\mu^{\alpha-1}
\ee
The ratio of the effective mass to the dominant frequency (in 
restored units of the square of inverse length) is directly proportional to 
$\alpha$, since
$m^*/F_0=2\pi\hbar^2\alpha/\mu$, and the absolute value of $m^*$ 
($m^*$ is negative in the case of $\alpha<0$) increases as
$\alpha$ increases when $F_0$ is kept fixed. This accounts for the
behavior of the oscillations reported in Fig.~\ref{fig2} whose amplitude become 
smaller as $\alpha$ increases. \cb{The maximum amplitude is obtained for the parabolic case}.

\subsection{Temperature dependence of the dominant frequency}
Here we consider the low temperature limit of the previous expressions
~\eref{AF_neg} and~\eref{AF_pos}, and evaluate the value of the dominant
frequency for $p=1$. At zero temperature, this corresponds to
$F_0=\mu^{\alpha}$, or to the first extremum of $A(F)$. We can 
perform an expansion around this value, by assuming the
parameter $\Delta=(F^{1/\alpha}-\mu)/2T$ to be small, $\Delta=O(T)$, and
extremizing the first harmonic term as function of $\Delta$. For both cases,
$\alpha$ negative and positive, one finds the following expansion
\bb\label{F0T}
F_0\simeq\mu^{\alpha}\left \{ 1+2\alpha(3-2\alpha)\frac{(k_BT)^2}{\mu^2} \right
\}
\ee
The coefficient in front of $(T/\mu)^2$ is always negative for $\alpha<0$ and
$\alpha>3/2$, and positive in the interval $0<\alpha<3/2$. This can be compared
to the fundamental frequency $F_0$ deduced from the numerical resolutions of 
Eqs.~\ref{AF_para} and~\ref{AF_neg} reported in the insets of~\efig{fig1}, 
where solid blue lines are low temperature approximations given 
by ~\eref{F0T} for $\alpha=1$, with the slope $2\alpha(3-2\alpha)=2$, and 
$\alpha=-1$, with the
slope $2\alpha(3-2\alpha)=-10$. Accordingly, the temperature dependence is 
steeper for $\alpha$=-1, while for $\alpha=(3-\sqrt{17})/4\simeq -0.28$ the 
slope ($-2$) is opposite to the one for the parabolic case.
%
\section{Low temperature expansion\label{section_LT}}
In this section, we consider the temperature behavior of the magnetization in
the low temperature limit $\beta\gg 1$. It is useful to consider the Fourier
transform of the function $\beta\de'$
\begin{align}\label{phi}
\beta\de'(\beta[\mu-E])=&\frac{\beta}{4\cosh^2\left
(\ff\beta [E-\mu]\right )}
=\beta\int_{ -\infty }^{ \infty}\frac{\drm g}{2\pi}
R(\pi g)\e^{i\beta g(E-\mu)}
\end{align}
which is convenient for a stationary phase approximation since the phase $\beta
gE$ of the exponential argument is large. Considering the case $\alpha>1$, the
magnetization~\eref{mag_alpha_pos} includes a triple integral over $g$, $u$ and
$E$
\begin{align}\nn
m_{osc}(b)=&\frac{2}{\alpha b}\Re\sum_{p\ge 1}
\e^{-2i\pi p\gamma}
\int_1^{\infty}\frac{\drm u}{u^{1-1/\alpha}}
\int_{ -\infty }^{ \infty}\frac{\beta \drm g}{2\pi}
R(\pi g)
\\ \label{RT}
\times&\int_0^{\infty}\drm E E^{1+\alpha}
\e^{2i\pi pE^{\alpha}u/b+ig\beta(\mu-E)}
\end{align}
We then apply the phase approximation to the argument function
$\varphi(E,g)=2i\pi pE^{\alpha}u/b+ig\beta(\mu-E)$ for which the stationary
solutions $\partial_E\varphi(E,g)=\partial_g\varphi(E,g)=0$ are given by
$E^*=\mu$ and $g^*=2\pi p\alpha u\mu^{\alpha-1}/\beta b$. One obtains a
temperature dependence of the magnetization in the general case as
\begin{align}\label{mag_alpha_pos_T}
m_{osc}(b)\simeq&\frac{2\mu^{1+\alpha}}{\alpha b}\Re\sum_{p\ge 1}
\e^{-2i\pi p\gamma}
\int_1^{\infty}\frac{\drm u}{u^{1-1/\alpha}}
R\left (\frac{2\pi^2p\mu^{\alpha-1}u}{\beta b} \right )
\e^{2i\pi p\mu^{\alpha}u/b}
\end{align}
which is valid only for $\alpha>1$. For negative values of $\alpha<0$, the same
analysis leads to the following approximation
\begin{align}\nn
m_{osc}(b)\simeq&-\frac{2\mu^{1+\alpha}}{b}\Re\sum_{p\ge 1}
\e^{-2i\pi p\gamma}
\left [
R\left (\frac{2\pi^2p\mu^{\alpha-1}}{\beta b} \right )
\e^{2i\pi p\mu^{\alpha}/b} \right .
\\ \label{mag_alpha_neg_T}
&\left . +\frac{2i\pi p\mu^{\alpha}}{b}
\int_1^{\infty}\frac{\drm u}{u^{-1/\alpha}}
R\left (\frac{2\pi^2p\mu^{\alpha-1}u}{\beta b} \right )
\e^{2i\pi p\mu^{\alpha}u/b}\right ].
\end{align}
The physical values of the temperature for which the oscillations can be seen
are bounded by the limit $\hbar\omega_c=k_BT$, or $\beta b=1$ in the reduced
units. This corresponds to the limit where the thermal energy is equal to the
Landau gap.
\section{Discussion and Conclusion}

The dHvA Fourier spectrum was analyzed for a class of Fermi surfaces whose area 
grows like $E^{\alpha}$,
by performing exact integrations of the oscillatory part of the grand 
potential harmonics.
This is relevant in particular for Dirac spectrum for which $\alpha=2$. 
All the deviations from parabolicity in the LK theory, which consists in 
a linearization of the Fermi surface, are incorporated in the main
formula~\eref{Ip}, from which the grand potential and the 
Fourier spectrum are calculated. The theory is not restricted to power law 
surfaces but can be extended to other type of energy dependence, for example in 
the case of singular surfaces with a strongly curvature dependence and where LK 
formula is not applicable. 
We can in particular extend the usual definition of the effective mass
$m^*=2\pi\hbar^2\partial S(\mu)/\partial \mu$ to all $\alpha$. In~\eref{SE},
$\alpha=1$ corresponds to
the effective mass of the quasiparticle, and in the second case the Dirac mass is given
by $m^*=\mu/v_F^2$, which is simply the Einstein
relation. As $\alpha$ increases, the effective mass increases as 
well, and the nature of the surface changes as well, which results in a reduction of the amplitude of the oscillations  as reported in~\efig{fig2}, irrespective of the temperature. 
The peaks of the Fourier spectrum are localized near the harmonics of the 
dominant frequency $F_0=\mu^{\alpha}$, and are temperature-dependent.
Whereas the dominant frequency of hole-type spectrum  decreases
as the temperature decreases, it increases for
electron-type spectrum, provided $\alpha$ is in the interval $0<\alpha<3/2$. 
Therefore, temperature dependence of the dominant frequency should also probe
the nature of the quasiparticles in the more general case. \cb{Electron-phonon interactions 
lead usually to an enhancement of the effective mass entering the LK damping factor which formula remains unchanged in the parabolic case~\cite{Sh84chap}. In general, the self-energy $\Sigma(E)$ for interactions needs to be incorporated in the quantization of the surface $S(E)\rightarrow S(E-\Sigma(E))$. At low temperature, below the Debye frequency, the self-energy can be approximated with a constant imaginary part, and we expect the effective mass to be renormalized in~\eref{eff_mass} by the same coefficient. Another question would be to determine how the amplitudes depend on the spin through the Land\'e factor, and how the usual spin-zero effect is affected. Indeed, a strong dependence of the amplitudes on the field direction is observed in experiments, while spin-orbit coupling only leads to a frequency splitting dependent on the field only, and not on temperature~\cite{Mineev:2005}. The spin-orbit coupling is treated as a field expansion of the dominant frequency around the extremal area. The dominant part of the extremal area is considered within a parabolic band model in this approach. The nature of the spin-zero effect is therefore a question which needs to be addressed in view of the temperature expansions~\eref{mag_alpha_pos_T} and~\eref{mag_alpha_neg_T}. Finally highly anisotropic materials~\cite{Goodrich:2004} could be investigated since they possibly may lead to non-parabolic behavior of the FS, due to the strong departure from its spherical shape in 3D compounds.}

\appendix
\section{Cosine Fourier transform}
It is relevant to consider the cosine Fourier transform of the magnetization
relatively to the inverse field when this one is taken positive $x=1/b>0$
\begin{align}
\nn
A(F)=&
4 \sum_{p\ge 1}p
\int_0^{\infty}x^2\cos(2\pi Fx)\drm x
\\ \nn
\times&\int_0^{\infty}\drm
EE^{1+2\alpha}\e^{2i\pi p(E^{\alpha}x-\gamma)}\beta
\de'(\beta[\mu-E])
\\ &\times
\int_0^{\infty}\drm u\e^{-puE^{\alpha}x}\left [
\left (1-\frac{u}{2i\pi}\right )^{1/\alpha}-1\right ]
\end{align}
from which amplitude the magnetization can be expressed as
$m_{osc}=2\Re \int_{0}^{\infty} \drm F A(F)\cos(2\pi F/b)$. After
integration
over $x$, one obtains
\begin{align}
\nn
&A(F)=
4\sum_{\sigma=\pm 1} \sum_{p\ge 1} \int_0^{\infty}\drm
E\frac{E^{1-\alpha}}{p^2}\e^{-2i\pi p\gamma}\beta
\de'(\beta[\mu-E])
\\ &\times \label{AF}
\int_0^{\infty}\drm u\left [ u+2i\pi\left (\frac{\sigma F}{pE^{\alpha}}-1\right
)
\right ]^{-3}
\left [
\left (1-\frac{u}{2i\pi}\right )^{1/\alpha}-1\right ]
\end{align}
The dependence in $F$ appears only in the product $F/pE^{\alpha}$ in the
last integrand and $A(-F)=A(F)$. In the parabolic case $\alpha=1$, the
amplitude is given by
\begin{align}
\nn
A(F)=&\frac{1}{\pi^2}\sum_{p\ge 1}(-1)^p
\int_0^{\infty}\drm E \beta \de'(\beta[\mu-E])\frac{E^2}{F^2-(p E)^2}
\\
=&\frac{1}{\pi^2}\int_0^{\infty}\drm E E^2\beta \de'(\beta[\mu-E])\frac{\pi
F/E-\sin(F/E)}{2F^2\sin(\pi F/E)}
\end{align}
The poles of this function are located at $F=\pm pE$, and the integration over
$F$ gives
\begin{align}\nn
m_{osc}&=2\Re \int_{0}^{\infty} \drm F A(F)\cos(2\pi F/b)
\\ \nn
=&\Re
\frac{1}{\pi^2}\sum_{p\ge 1}(-1)^p
\int_0^{\infty}E^2\drm E \beta \de'(\beta[\mu-E])
\int_{-\infty}^{\infty} \drm F
\frac{\e^{2i\pi F/b}}{F^2-(p E)^2}
\\ \nn
=&\Re
\frac{1}{2\pi^2}\sum_{p\ge 1}\frac{(-1)^p}{p}
\int_0^{\infty}E\drm E \beta \de'(\beta[\mu-E])
\int_{-\infty}^{\infty} \drm F \e^{2i\pi F/b} \left (
\frac{1}{F-p E}-\frac{1}{F+p E} \right )
\\
=&-\frac{1}{\pi}\sum_{p\ge 1}\frac{(-1)^p}{p}
\int_0^{\infty}E\drm E \beta \de'(\beta[\mu-E])
\sin(2\pi pE/b)
\end{align}
This formula just corresponds to~\eref{mosc_para} derived in 
section 2.


\section*{References}
\bibliography{curvature}

\end{document}